\documentclass[journal=jacsat,manuscript=article]{achemso}

\usepackage{booktabs}
\usepackage{bm}

\usepackage{comment}
\usepackage{tabularx}
\usepackage{color}
\usepackage{xcolor}
\usepackage{wrapfig}
\usepackage{float}
\usepackage{appendix}
\usepackage{bbm}

\usepackage{mathrsfs}
\usepackage{amsmath, amssymb, amsfonts, amsthm,rotating,booktabs,array}
\usepackage{cancel}
\usepackage{bm}
\usepackage{booktabs}
\usepackage{subfigure}
\usepackage{graphicx}
\usepackage{mathabx}
\usepackage{setspace}
\usepackage{arydshln} 
\usepackage{algorithm,algpseudocode}
\usepackage{mathtools}
\usepackage{multicol}
\usepackage{multirow}
\usepackage{verbatim}
\usepackage{enumerate}

\usepackage[version=3]{mhchem} 

\author{Saejin Oh}
\altaffiliation{These authors contributed equally to this work.}
\affiliation{BioPACIFIC Materials Innovation Platform, University of California, Santa Barbara, Santa Barbara, CA, 93106 USA}

\author{Xinyi Fang}
\altaffiliation{These authors contributed equally to this work.}
\affiliation{Department of Statistics and Applied Probability, University of California, Santa Barbara, Santa Barbara, CA, 93106 USA}

\author{I-Hsin Lin}
\affiliation{Department of Materials Science and Engineering, University of California, Irvine, CA, 92697 USA}

\author{Paris Dee}
\affiliation{Department of Chemistry and Biochemistry, University of California, Los Angeles, Los Angeles, CA, 90095 USA}

\author{Christopher S. Dunham}
\affiliation{BioPACIFIC Materials Innovation Platform, University of California, Santa Barbara, Santa Barbara, CA, 93106 USA}

\author{Stacy M. Copp}
\affiliation{Department of Materials Science and Engineering, University of California, Irvine, CA, 92697 USA}
\alsoaffiliation{Department of Chemical and Biomolecular Engineering, University of California, Irvine, CA, 92697 USA}
\alsoaffiliation{Department of Physics and Astronomy, University of California, Irvine, CA, 92697 USA}
\alsoaffiliation{Department of Chemistry, University of California, Irvine, CA, 92697 USA}
\email{copps@uci.edu}

\author{Abigail G. Doyle}
\affiliation{Department of Chemistry and Biochemistry, University of California, Los Angeles, Los Angeles, CA, 90095 USA}
\email{abigaildoyle@g.ucla.edu }

\author{Javier Read de Alaniz}
\affiliation{Department of Chemistry and Biochemistry, University of California, Santa Barbara, Santa Barbara, CA, 93106 USA}
\email{jalaniz@ucsb.edu}

\author{Mengyang Gu}
\affiliation{Department of Statistics and Applied Probability, University of California, Santa Barbara, Santa Barbara, CA, 93106 USA}
\email{mengyang@pstat.ucsb.edu}
\title[automated lab]
{Synergizing chemical and AI communities for advancing laboratories of the future}

\abbreviations{IR,NMR,UV}

\begin{document}

\begin{tocentry}
    \centering
    \includegraphics[width=\linewidth]{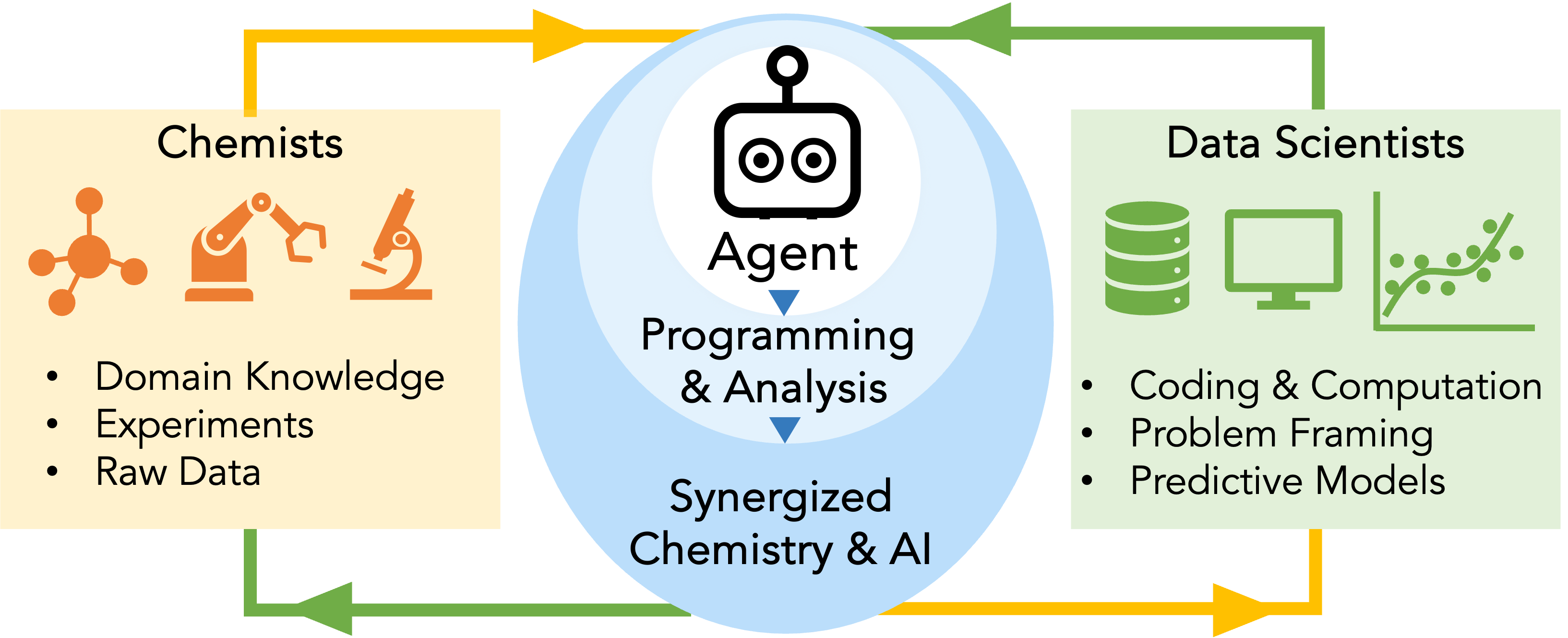}
\end{tocentry}

\newpage
\begin{abstract}
The development of automated experimental facilities and the digitization of experimental data have introduced numerous opportunities to radically advance chemical laboratories. As many laboratory tasks involve predicting and understanding previously unknown chemical relationships, machine learning (ML) approaches trained on experimental data can substantially accelerate the conventional design-build-test-learn process. This outlook article aims to help chemists understand and begin to adopt ML predictive models for a variety of laboratory tasks, including experimental design, synthesis optimization, and materials characterization. Furthermore, this article introduces how artificial intelligence (AI) agents based on large language models can help researchers acquire background knowledge in chemical or data science and accelerate various aspects of the discovery process. We present three case studies in distinct areas to illustrate how ML models and AI agents can be leveraged to reduce time-consuming experiments and manual data analysis. Finally, we highlight existing challenges that require continued synergistic effort from both experimental and computational communities to address.    
\end{abstract}


\section{Introduction}

 Laboratory experiments are one of the most critical conduits to advance basic science and technology. In recent years, the field of chemistry has experienced numerous significant milestones in accelerating laboratory experiments with the introduction of critical techniques, including robotic arms, computational facilities, machine learning (ML) algorithms, and artificial intelligence (AI) agents based on large language models (LLMs). These advancements automate various laboratory processes, ranging from synthesis and purification to characterization and data analysis with minimal human intervention, 
 stimulating the transition towards self-driving laboratories 
\cite{abolhasani2023rise, tom2024self, vriza2023self, wang2025autonomous, seifrid2022autonomous, macleod2020self, zhang2024leveraging, wu2023materials}.

\begin{figure}[t]
    \centering
    \includegraphics[width=\linewidth]{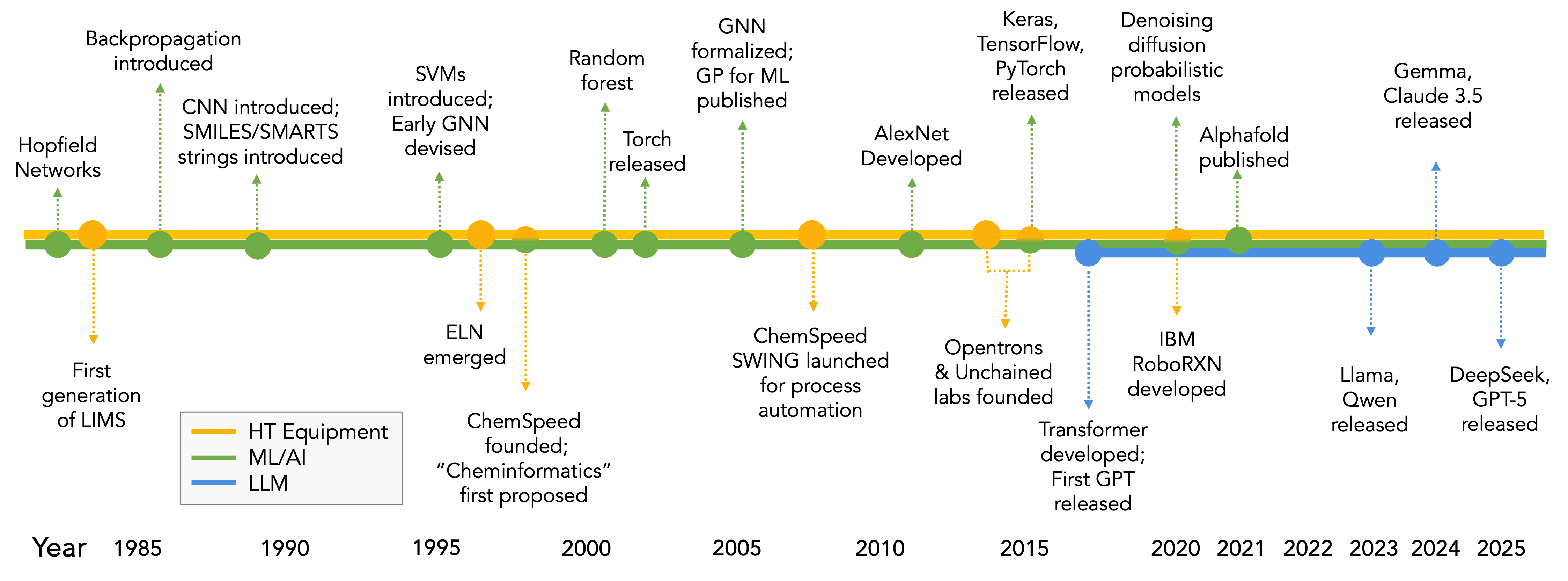}
    \caption{A brief timeline for the major developmental milestones of HT equipment, ML/AI algorithms, and LLMs for the labs of the future.}
    \label{fig:Timeline_revised}
\end{figure}

 Figure \ref{fig:Timeline_revised} shows a timeline of the introduction of selective high-throughput (HT) experimental facilities, ML/AI algorithms, and LLMs over the past three decades. Although  automated and self-driving laboratories are a relatively new concept, 
 tools for tracking and cataloging data for experimentation, such as laboratory information management systems (LIMS) \cite{gibbon1996brief} and electronic laboratory notebooks (ELNs) \cite{du2007electronic, rubacha2011review}, 
  were conceptualized 30-40 years ago. As data acquisition and processing became 
  increasingly multi-step and time-consuming, automated and parallel operations of HT experiments have evolved in different areas.\cite{tom2024self, dai2024autonomous, ha2023ai} For example, Chemspeed, one of the largest lab automation hardware companies for chemical synthesis, was founded in the late 1990s and introduced key products such as the SWING platform in 2007, which 
 enabled automated formulation screening in a high-throughput way. 
 As another example, Unchained Labs 
 was founded in 2015 and launched various automated instruments dedicated to bio-applications.

 The hardware of laboratory research has evolved along with the 
 computational tools capable of powering the feedback loops that guide operations. For instance, algorithms, such as backpropagation \cite{rumelhart1985learning}, one of the most useful approaches to optimize artificial neural networks,\cite{hopfield1982neural} were formally introduced in the early 1980s. The 1990s and early 2000s saw the development of ensemble tree techniques, such as random forests, and probabilistic models, including Gaussian processes,
for nonlinear regression and classification problems with small to moderate data sizes.\cite{breiman1996bagging,breiman2001random,friedman2001greedy,rasmussen2006gaussian} With the arrival of massive data collections of text and images on the Internet, different architectures of neural networks, such as convolutional neural networks and recurrent neural networks, were developed and evolved to be more flexible and accurate for tasks such as image classification and segmentation.\cite{hochreiter1997long,lecun2002gradient,ronneberger2015u}  The development of neural network architectures \cite{hopfield1982neural, ackley1985learning} and their profound impacts in predicting protein structures \cite{jumper2021highly,baek2021accurate,abramson2024accurate} was awarded the 2024 Nobel Prizes in Physics and Chemistry, respectively. Trained by simulated or experimental data,  ML  methods can be routinely used as models 
for predicting untested inputs,\cite{ahneman2018predicting,deringer2021gaussian} which can facilitate operations in almost all areas of laboratory science, including experimental design, synthesis optimization, and materials characterization.\cite{zhang2020bayesian,shields2021bayesian}

Over the past decade, generative AI models based on transformer architecture \cite{vaswani2017attention} and score-based generative models \cite{ho2020denoising,song2020score}
have gained tremendous attention across the world for text and image generation, and have opened up a new era of scientific research. 
The transformer, a neural network architecture for training LLMs, for instance, 
 inspired the development of  
the Generative Pre-trained Transformer (GPT),\cite{radford2018improving,achiam2023gpt} and other LLM models, such as Claude, Gemini,  Llama, Qwen, and DeepSeek.\cite{naveed2023comprehensive, touvron2023llama,yang2025qwen3,liu2024deepseek,guo2025deepseek} 
The versatility of LLMs for use in a variety of operations, ranging from literature summary to computer code generation, reduces barriers to learning new disciplines and facilitates interdisciplinary collaboration, which has started to transform the paradigm in chemical laboratory research. \cite{boiko2023autonomous,ramos2025review} Furthermore, score-based generative models, such as denoising diffusion models, have been applied for protein structure prediction and design.  \cite{watson2023novo,abramson2024accurate}

\begin{figure}[t]
    \centering
    \includegraphics[width=\linewidth]{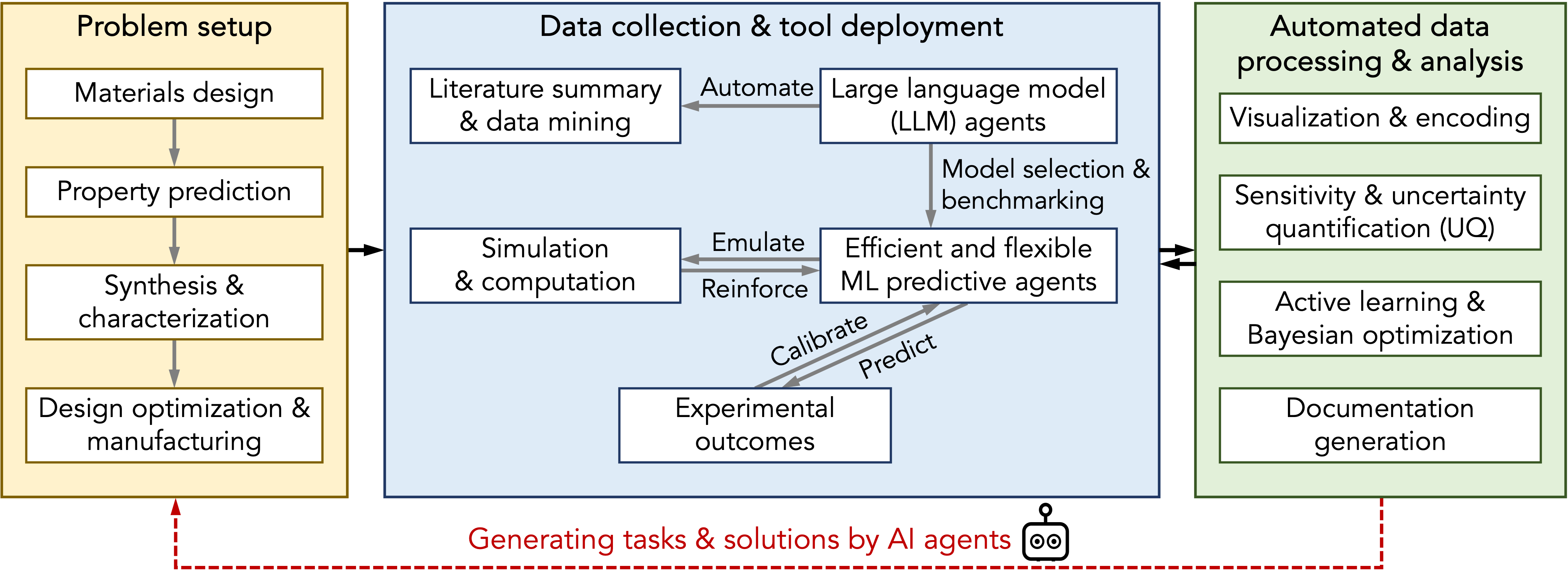}
    \caption{Laboratory workflows automated and accelerated by agentic AI.}
    \label{fig:workflow}
\end{figure}

Today, we stand at a pivotal moment for radically transforming laboratory research and education. Traditional chemical laboratories require significant human labor for manual experimental designs, product screening, and data analysis, which can be substantially accelerated by robotic systems and AI agents,  illustrated by the workflows in Figure \ref{fig:workflow}. 
The  LLMs and ML predictive models can encode multiscale, cross-disciplinary information, enabling scalable and accurate prediction for a large number of test samples,  thereby substantially reducing the experimental cost and time. 
 However, many researchers, particularly in experimental science, are unsure where to begin and what ML methods they should use to minimize deployment effort and cost.  
 Although research tasks can be drastically different between chemical science communities, many involve forming, predicting, and understanding chemical relationships, i.e.  $f: \mathbf x \rightarrow f(\mathbf x)$, where $\mathbf x$ can be descriptors of molecules, chemicals, experimental conditions or experimental outcomes, such as microscopy images and scattering curves, and $f$ is a function that maps the input to system properties, such as conductivity, chemical reaction yields,  structural and mechanical properties of the materials. Our modern world is built upon the discovery of maps that accurately predict previously unknown relationships. In the past, however, to discover the underlying principles 
 of a new system, chemists often relied on time-consuming lab experiments and manual analysis of data in a traditional lab. 

Two critical advances have paved the way for data-driven discovery of unknown relationships in chemical science. First, experimental and simulation data have gradually become digitalized, enabling the use of fundamental statistical learning principles, such as Bayes' theorem, to automatically update rules from the \textit{status quo}, or prior distribution, to a new paradigm, or posterior distribution, by conditioning on new data. Second, ML models have advanced over the years to learn complex relationships from data, such as numerical values, texts, and sequences, which can substantially reduce time and computational cost for analyzing complex data. Through the lens of these changes, this outlook article will assess the current status of chemical laboratory research, highlight existing gaps, and suggest a path for uniting experimental and computational communities to accelerate progress.

\section{Accelerating Data Collection and Processing}

\begin{figure}[t]
    \centering
    \includegraphics[width=\linewidth]{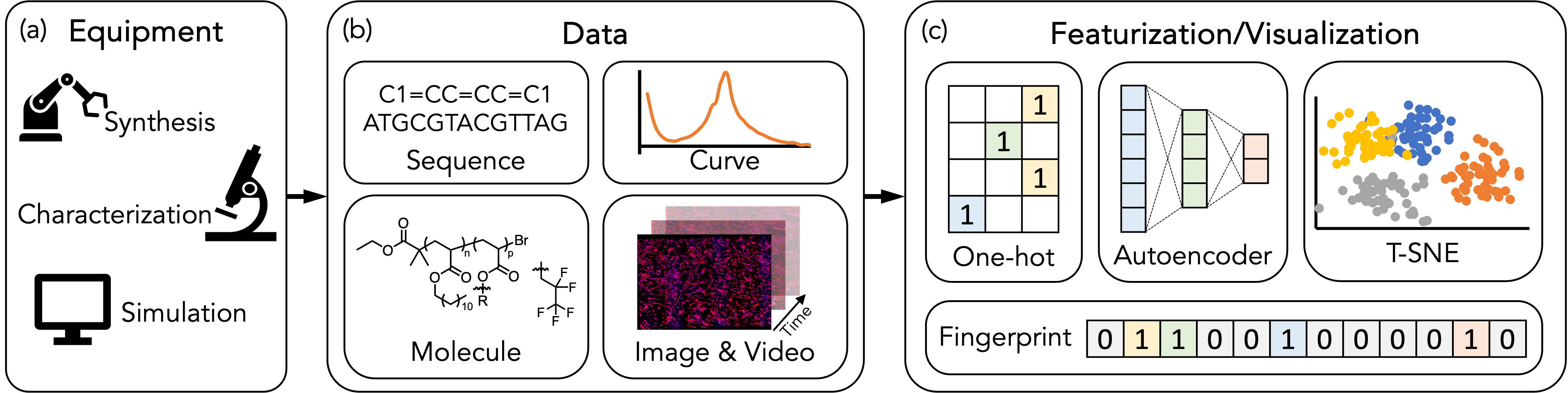}
    \caption{Data collection, processing and featurization in chemical research. }
    \label{fig:Featurization}
\end{figure}

{\bf Data Acquisition.} 
Materials synthesis, characterization, and simulation are three main sources of chemical data, shown in Fig.  \ref{fig:Featurization}(a), which produce, for instance, 
molecular sequences, curves, images, and videos  (Fig. \ref{fig:Featurization}(b)). The key goals 
are to accelerate and automate data collection, processing, and featurization (Fig. \ref{fig:Featurization}(c)) for guiding the process of learning chemical relationships. 

First, advances in automation are transforming the way materials are synthesized and fabricated for downstream analysis.\cite{caramelli2021discovering, ha2023ai, tom2024self} Robotic platforms 
can be flexibly programmed to perform a range of chemical reactions and formulations with high precision and reproducibility, enabling parallel experimentation in multi-well plate formats.\cite{seifrid2024chemspyd, seifrid2022autonomous, tom2024self}  Flow chemistry further extends automation by providing continuous control over reaction conditions, incorporating in-line characterization tools for real-time monitoring, and improving safety when handling hazardous compounds.\cite{plutschack2017hitchhiker, guidi2020approach, capaldo2023field}
Once reactions are complete, automated flash purification systems and preparative high-performance liquid chromatography \cite{still1978rapid} streamline isolation of small molecules and can be adapted to generate well-defined polymer libraries with minimal human intervention.
\cite{murphy2024chromatographic, fang2025universal}. Beyond producing physical samples, these automated platforms generate distinct types of  records, including molecular structures, reaction conditions, and experimental procedures, which can be digitized into machine-compatible formats. For instance, information on molecular structures can be converted into SMILES and SELFIES strings.     \cite{weininger1988smiles,lin2019bigsmiles, krenn2022selfies}  Furthermore, efforts are being made to standardize experimental procedures, such as the Open Reaction Database \cite{kearnes2021open} and Chemical Description Language \cite{mehr2020universal}, for training ML models to optimize synthesis and reaction conditions. 
Commonly used methods to represent discrete inputs include one-hot encoding, which expresses discrete inputs by sequences of `0' and `1', and molecular fingerprints by numerical vectors  \cite{todeschini2008handbook,rogers2010extended,sandfort2020structure}. 
Encoding these methods helps bridge synthesis outputs with machine learning models that can analyze reaction trends and accelerate discovery. 

Second, a wide range of materials characterization tools, including microscopy,  rheology, spectrometry, scattering, and spectroscopy, have been developed. These tools generate images, time-series data, spectra, or other quantitative values in chemical laboratories. Data processing tools, such as image segmentation and particle tracking,  \cite{crocker1996methods} have been developed for extracting and linking data from microscopy images. These data processing tools have been implemented into software packages, such as ImageJ and  Fiji  \cite{schneider2012nih,schindelin2012fiji}, which contain easy-to-use graphical user interfaces (GUIs), empowering users to view and analyze large quantities of data, particularly useful for biochemical research.\cite{luo2023cell}  
The availability of a high volume of labeled data enables the development of more accurate supervised learning tools, such as Cellpose \cite{stringer2021cellpose}, which utilizes a large database of labeled data to train U-Net \cite{ronneberger2015u}, a convolutional neural network for segmenting cells from microscopy images. For more challenging scenarios, such as capturing optically dense systems and fast dynamics, Fourier-based tools, e.g. differential dynamic microscopy (DDM) \cite{cerbino2008differential,giavazzi2009scattering}, remove the need to segment particles to extract system information, e.g. mean squared displacement of the particles, that determine the mechanical properties (storage, loss modulus) \cite{bayles2017probe,gu2021uncertainty}. Building upon existing tools, it is possible to construct probabilistic generative models 
and automated estimators for  existing data processing methods, such as 
by removing 
manual selection of the Fourier range in DDM \cite{gu2024ab} which otherwise needs to be chosen on a case-by-case manner \cite{martinez2012differential,lu2012characterizing,bayles2016dark,guidolin2023protein}.

Third, computational simulations from distinct space-time length scales can provide scientific insights and a pathway to explore chemical systems before conducting chemical experiments.\cite{RN45,rapaport2004art,fredrickson2006equilibrium}. These simulations can reveal mechanistic insights prior to experimentation but are often limited by large computational and/or storage costs, and the need for accurate model calibration, such as determining the form of observed model parameters.\cite{sholl2022density, koch2015chemist, matsen2006self} 
To address this challenge, Meta FAIR has released Open Molecules 2025 (OMol25), a large-scale open-source dataset comprising over 100 million density functional theory (DFT) calculations. It aims to provide high-accuracy quantum chemical data to support the development of machine learning models in molecular chemistry.
\cite{levine2025openmolecules2025omol25}
The past decade witnessed the success of ML surrogate models \cite{rupp2012fast,bartok2013representing,brockherde2017bypassing,chmiela2017machine,li2022efficient,lu202186,behler2021four,sammuller2023neural} for predicting outcomes of expensive simulations, such as the potential energy, force field, and particle density at untested inputs from nanoscale to bulk environment. For example, neural network potentials and Gaussian process regression have been used to accelerate molecular dynamics and DFT calculations.\cite{smith2017ani, duignan2024potential, deringer2021gaussian} Integrating ML-accelerated simulations into laboratory workflows can reduce the number of experiments in labs and guide synthesis toward the most promising targets. Realizing this vision requires closer collaboration between experimental and computational communities, ensuring that simulation-informed predictions are seamlessly incorporated into automated experimentation and data-driven discovery workflows.

As the tools used to inform laboratory operations have expanded and evolved, so has the need to record and manage data from these systems. Software, such as LIMS and ELNs, is capable of providing mechanisms for researchers to catalog and record key experimental data in ways that are searchable, labeled, uniquely identified, and accessible in machine-readable formats. Additionally, digital representations of laboratory protocols and associated data can simplify sharing and enable greater collaboration between researchers. The information in an ELN can be utilized to provide training data to update data-driven methods for prediction and optimization. Because of these advantages,  physical notebooks of laboratories are gradually being replaced by ELNs.\cite{rubacha2011review, machina2013electronic} 
Furthermore, data from an ELN can be stored in or connected to a LIMS to enable comprehensive lab data management.\cite{paszko2018laboratory, skobelev2011laboratory, prasad2012trends} Together, ELN and LIMS serve as tools that can foster open access data for researchers to retrieve, review, and analyze.

\textbf{Input Featurization and Visualization.} As the input or descriptor $\mathbf x$ is not often available to learn chemical relationships $f(\mathbf x)$,  domain knowledge, cheminformatics, and simulation are often used to generate feature sets that capture underlying chemical structures. Representative cheminformatics packages, including OpenBabel, RDKit, and CDK, have been  integrated with popular programming languages (Table \ref{tab:cheminformatics_packages}), \cite{o2008cinfony} 
which enables processing scientific data to obtain meaningful input features for a wide range of problems.  

\begin{table}[t]
    \centering
    \begin{tabular}{l l l}
        \toprule
        Cheminformatics Package & Languages & Strength \\ 
        \midrule
        OpenBabel & C++, Python, Java & Format conversion, Structure search \\
        RDKit     & C++, Python       & Molecular analysis, ML \\
        CDK       & Java              & Computational chemistry, Bioinformatics \\
        \bottomrule
    \end{tabular}
    \caption{Examples of typical cheminformatics packages.}
    \label{tab:cheminformatics_packages}
\end{table}

Furthermore, exploratory data analysis tools are commonly used for visualization and featurization \cite{tukey1977exploratory}. A common challenging scenario for featurization involves high-dimensional data, including curves, images, or videos,
and discrete inputs such as molecular sequences and graphs. 
   Unsupervised dimension reduction tools, such as principal component analysis \cite{jolliffe2002principal}, t-distributed stochastic neighbor embedding (t-SNE) \cite{van2008visualizing}, uniform manifold projection and reduction (Umap) \cite{mcinnes2018umap},  dynamic mode decomposition \cite{schmid2010dynamic}, autoencoders and decoders \cite{kingma2013auto}, are developed for extracting features of high-dimensional data.  
These methods can be used to visualize the high-dimensional datasets, and the reduced dimensionality vectors can be input as features for ML models. 
Domain knowledge, such as physical and chemical principles, can also be used to reduce the dimension of data and improve the accuracy of noisy experimental data. For instance, for classifying phases of block copolymers by small-angle X-ray scattering (SAXS) data, using several features relevant to the location, width, and curvature of the primary peaks of the X-ray curves substantially improves the predictive accuracy of ML models compared to using the entire curve as input in ML models \cite{fang2025universal}. Furthermore, scattering measurements were used to estimate the micelle structure of block copolymer solutions inversely,\cite{beltran2019computational} and ML surrogate models can improve the inverse estimation by learning the map from reduced-dimensional features of micelle structural parameters to scattering patterns.\cite{heil2022computational}

Another common challenge of featurization involves discrete or categorical inputs, such as different types of atoms, molecules, and  chemical bonds.  The overarching goal of featurization 
 is to inform the ordering of chemical candidates in terms of their system properties. Compared with numerical inputs, discrete inputs are more challenging to model due to the lack of ordering between the inputs.  ML models have achieved success for predicting discrete sequences in some applications, including transformers in LLMs that predict the next text token given the context  \cite{vaswani2017attention}, and AlphaFold that maps amino acids to protein spatial structure. \cite{jumper2021highly}  These examples demonstrate the importance of standardized data sets and novel ML architectures for modeling discrete inputs.

\section{Learning Chemical Relationships by Predictive Models}

\begin{figure}[t]
    \centering
    \includegraphics[width=\linewidth]{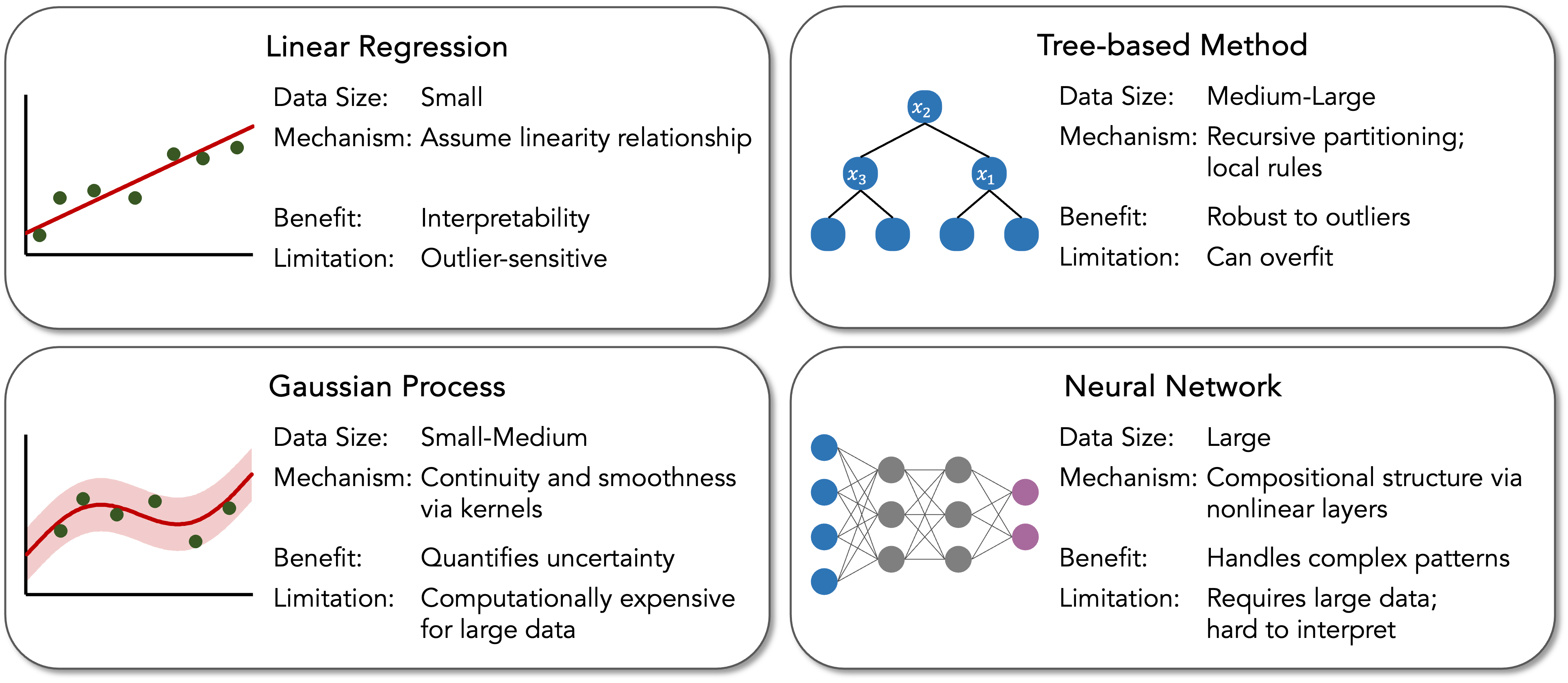}
    \caption{Data-driven predictive models for chemical research.}
    \label{fig:ML_chemical_research}
\end{figure}

\textbf{Predictive Models.}  A predictive model, sometimes referred to as statistical methods of chemometrics by chemists,\cite{williams2021evolution} is an indispensable component for learning chemical relationships. 
With a given input vector $\mathbf x$, a common goal is to predict the function $f(\mathbf x)$ that maps the input to system properties, 
and quantify the uncertainty of the prediction. Such a process typically involves training a data-driven predictive model and making predictions. We will first start from predicting real-valued outcomes, which is generally known as the regression, and introduce 4 classes of widely used predictive models, listed in Figure \ref{fig:ML_chemical_research}. All these models can be generalized to predict categorical data and counts, generally known as classification, by defining a link function, such as the logistic function \cite{hastie2009elements}, to map the numerical outcomes to the probability of each categorical outcome.

A linear model is potentially the oldest and most widely used benchmark model. Assume the input $\mathbf x$ is a vector of $p$ variables, $\mathbf{x} = [x_1, x_2, ..., x_p]^T$. 
The model assumes the relationship is linear $f(\mathbf x) = \beta_0 + \beta_1x_1 + \dots+\beta_px_p$, where $\bm \beta=[\beta_0,...,\beta_p]^T$ is a vector of  coefficients to be estimated from data. 
Statistical theory has been well established for estimating the coefficient of linear regression for noisy observations.  
Due to the assumption of linearity, linear models typically do not require large amounts of data to estimate the parameters.  With the use of shrinkage methods \cite{tibshirani1996regression,efron2004least}  
that penalize large coefficients, the number of observations can be much smaller than the number of variables in the system. These shrinkage estimators avoid exploring the massive variable space needed to solve computationally expensive combinatorial problems, which have found applications, for instance, in discovering mathematical equations \cite{brunton2016discovering}. 
In addition to prediction, linear methods offer a rigorous framework for statistical inference, hypothesis testing, and variable selection for automating model construction.\cite{luo2022high,gu2023data} Therefore, though the predictive power of a linear model is constrained by its restrictive assumption, the interpretability and the ease of fitting the linear model make it a suitable benchmark model to estimate unknown chemical relationships.

Tree-based ensemble methods,\cite{breiman2017classification} such as random forests \cite{breiman2001random,liaw2002classification} and gradient-boosted trees \cite{friedman2001greedy}, generalize the linear models by assuming locally linear relationships through partitioning the variable or feature space. They are widely used for their robustness and ability to model nonlinear relationships. 
Random forests, for instance, construct multiple decision trees in parallel, each trained on a bootstrap sample and a randomly selected subset of features. Predictions are obtained by aggregating across all trees, via the majority vote for classification or averaging for regression, thus reducing variance and mitigating overfitting. 
In contrast, gradient-boosted trees are built sequentially, with each new tree focusing on correcting the residuals or errors of the previous model. These methods naturally handle both numerical and categorical inputs, are insensitive to feature scaling, and are computationally efficient. In addition, they provide feature importance metrics based on the reduction of impurity or gain in predictive power at each split. This allows researchers to identify key structural features that dominate the properties of molecules or materials. 

Gaussian process regression is a flexible, nonparametric
approach for modeling nonlinear relationships and quantifying uncertainty in predictions\cite{rasmussen2006gaussian}.  For a continuous function with either scalar or vectorized outputs \cite{gu2016parallel}, the outcome values become more similar or more correlated when corresponding inputs become closer, which can be modeled by a kernel function in a Gaussian process. 
Conditioning on a set of observations, the predictive distribution of Gaussian process regression provides both predictions and uncertainty quantification. Compared to linear models and tree-based models, Gaussian processes are more efficient to learn nonlinear relationships, and often less training data is needed when the underlying map is smooth.  When the sample size is large, approximation methods \cite{snelson2006sparse,vecchia1988estimation} are often required due to the computational expense for Gaussian processes. The high efficiency with respect to small samples and availability of uncertainty make the Gaussian process a suitable candidate for surrogate models in predictions and design optimization \cite{deringer2021gaussian}.

Artificial neural networks 
are capable of learning intricate patterns from large datasets. A feedforward neural network is mathematically formulated as a composition of nonlinear functions $f(\mathbf x)=f^{(L)}(f^{(L-1)}(\dots f^{(1)}(\mathbf{x})))$, where each layer function $f^{(l)}(\mathbf{x}^{(l-1)}) = \sigma(\mathbf W^{(l)}\mathbf{x}^{(l-1)} + \mathbf{b}^{(l)})$ consists of a weight matrix $\mathbf W^{(l)}$, a bias vector $\mathbf{b}^{(l)}$, a nonlinear activation function $\sigma(\cdot)$ that acts element-wise on each coordinate of input vector, with the input at the first layer denoted by $\mathbf x^{(0)}=\mathbf x$. The large number of parameters enables neural networks to effectively learn a latent input space when the correlation between the outputs is hard to model. 
In recent years,  many neural network architectures \cite{lecun2015deep}, such as convolutional neural networks \cite{lecun2002gradient} and recurrent neural networks \cite{hochreiter1997long}, have found great success particularly for image analysis such as image classification \cite{krizhevsky2012imagenet}, segmentation \cite{ronneberger2015u}, generation and inpainting \cite{ho2020denoising,song2020score}. 
As the neural network models often require a large amount of data to train, they are suitable for certain scenarios 
such as learning  potential energy and atomic forces from simulation \cite{behler2007generalized,zhang2018deep}, and segmenting cells from microscopy images \cite{stringer2021cellpose}.

\begin{table}[t]
    \centering
    \begin{tabular}{l l l}
        \toprule
        Predictive Models & Python Packages & R Packages \\ 
        \midrule
        Linear regression & scikit-learn\cite{pedregosa2011scikit} & stats\cite{Rbase}, glmnet\cite{friedman2010regularization} \\
        Tree-based models & scikit-learn, XGBoost\cite{chen2016xgboost} & randomForest,\cite{liaw2002classification} xgboost\cite{xgboostR} \\
        Gaussian processes & scikit-learn, GPyTorch\cite{gardner2018gpytorch} & RobustGaSP\cite{gu2018robustgasp}, GpGp\cite{GpGpR} \\
         Neural networks & PyTorch\cite{paszke2019pytorch}, TensorFlow\cite{abadi2016tensorflow}, Keras\cite{chollet2015keras} & torch\cite{torchR}, keras\cite{chollet2017kerasR}\\
        \bottomrule
    \end{tabular}
    \caption{Examples of Python and R packages for predictive models.}
    \label{tab:ML_packages}
\end{table}

 Examples of the Python and R packages for the four classes of predictive models are given in Table \ref{tab:ML_packages}. These approaches have been widely used for predicting experimental outcomes \cite{ahneman2018predicting,luo2025optimizing} or as a surrogate model for approximating computationally expensive simulations \cite{chmiela2017machine}.   
 In practice, it is also critical to have reliable uncertainty quantification of the predictions, expressed as predictive intervals, for optimizing experimental designs \cite{snoek2012practical} and controlling predictive error \cite{fang2022reliable}.   
As linear regression and Gaussian processes are probabilistic models, the uncertainty of the predictions can be naturally expressed by predictive intervals based on the probabilistic framework. The uncertainty of Bayesian additive tree methods can be obtained from posterior samples \cite{chipman2010bart}, and quantile regression methods and asymptotic analysis were developed for quantifying the uncertainty of the ensemble tree methods \cite{meinshausen2006quantile,mentch2016quantifying}.  Assessing the uncertainty of neural network approaches is still an open area of research, and various methods, such as dropout, ensemble samples, and conformal estimation, were developed to quantify the sensitivity and uncertainty of neural networks \cite{srivastava2014dropout,gal2016dropout,lakshminarayanan2017simple,kristiadi2020being,fontana2023conformal}.  

\textbf{Experimental design optimization.} 
Leveraging the predictive power from simulation and ML methods enables the efficient design of experiments to understand an enormous space of molecules and materials. A primary goal of efficient materials design can be mathematically formulated as an optimization problem: $\mathbf{x}^* = \arg\max_{\mathbf{x}} g(\mathbf{x})$, where 
$g(\mathbf x)$ is the gain function of system properties from experimental outcomes under given input $\mathbf x$ (such as materials and experimental conditions). The challenge here is that the objective function $g$ is usually a ``black box" function that contains experimental noises, and the enormous input design space, which prohibits conducting experiments for each input point. 
 Applying traditional optimization methods such as quasi-Newton's method \cite{nocedal1999numerical} typically requires gradient information, noise-free outcomes of the objective functions, and a relatively large number of evaluations. 
To overcome these challenges, a predictive model, such as a Gaussian process, can be used as a probabilistic proxy to sequentially design the next experiments that give the most valuable experimental outcome through an acquisition function, a process often referred to as Bayesian optimization or active learning \cite{frazier2018tutorial}. The quantified uncertainty from the predictions is crucial to strike a balance between exploration and exploitation for making better predictions and improving the gain function, respectively \cite{wang2023recent}.

\begin{figure}[t]
    \centering    \includegraphics[width=\linewidth]{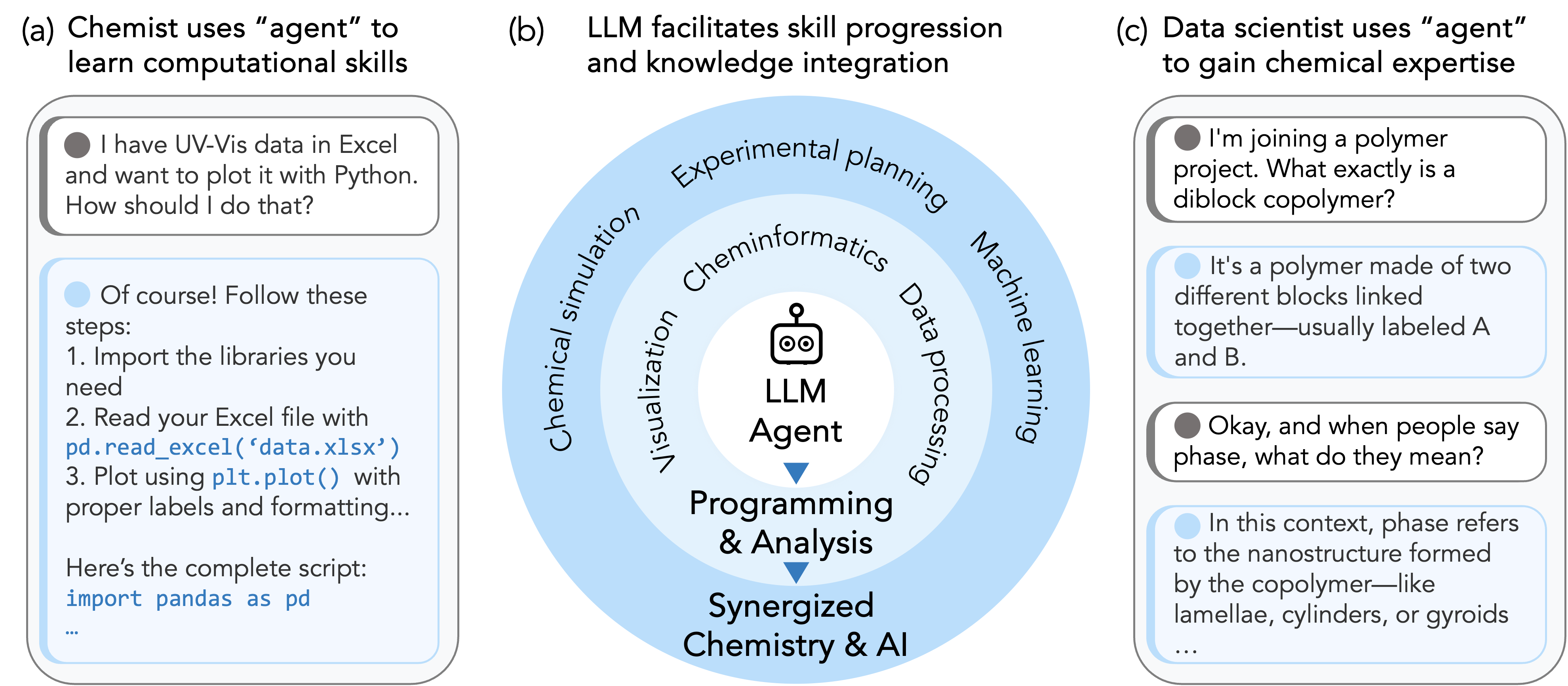}
    \caption{LLM agents facilitate cross-disciplinary collaboration and skill development in chemical research. (a) Example dialogue of chemists acquiring Python programming skills for data analysis. (b) Skill progression framework from basic computational tools to advanced chemistry-AI applications. (c) Example dialogue of LLM agents helping explain chemical concepts.}
\label{fig:LLM_agent_dialogue}
\end{figure}
\section{Filling the Gaps by LLM Agents}

Advancing laboratory research involves a large set of tools and techniques. Thus, it is imperative to educate students and researchers on the evolving approaches in automated facilities and data science, and  framing the laboratory research tasks as properly defined mathematical problems for data scientists. 

The rise of LLMs, such as ChatGPT, offers a promising path forward in connecting distinct domains to   accelerate learning and problem formulation processes, where the LLMs act as the agent at the interface between chemists and data scientists. 
Figure \ref{fig:LLM_agent_dialogue} illustrates several potential applications of LLMs, including generating computer code to perform data analysis for chemists and helping computational experts better understand concepts in chemistry. By accelerating learning processes and reducing communication barriers, LLMs can serve as helpful mediators to facilitate collaborations between distinct communities.

Several recent studies have explored the use of LLMs in chemical research, including assisting with coding and framing scientific questions using chemical data. \cite{white2023future, jablonka202314,griffen2020chemists} LLMs offer an accessible entry point for novices lacking computational skills, enabling efficient data processing, high-quality visualization, \cite{subasinghe2025large} 
and generating computer codes 
with only minimum prior programming experience. \cite{hare2024coding, nam2024using} 
In surveys conducted after introducing LLMs as learning tools, users reported notable improvements in their coding skills, demonstrating that LLMs can accelerate learning with minimal barriers. \cite{hare2024coding} Beyond basic use, LLMs can support general chemistry problem-solving \cite{white2023assessment}, and they can be fine-tuned for domain-specific tasks to further enhance output quality. \cite{jablonka2024leveraging}

\begin{figure}[t]
    \centering
\includegraphics[width=0.9\linewidth]{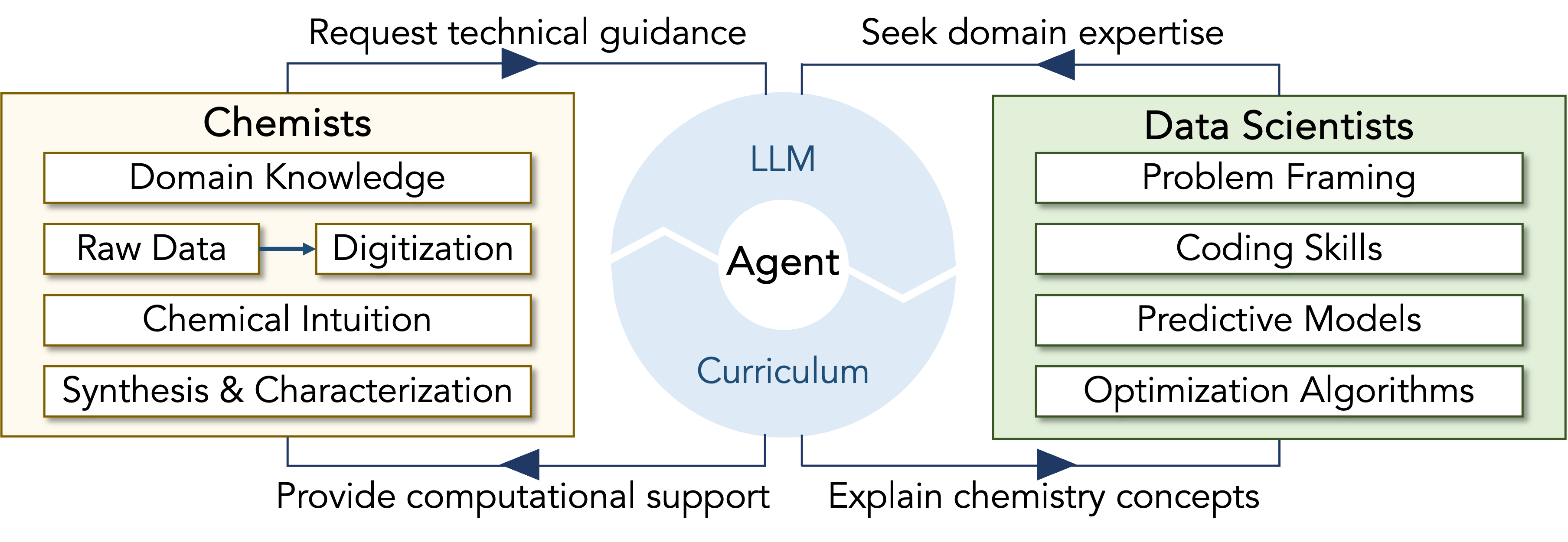} \caption{Collaborative workflows between chemists and data scientists facilitated by LLM agents.}
\label{fig:fill_gap_workflow}
\end{figure}

Figure \ref{fig:fill_gap_workflow} provides examples of distinct expertise from chemists and data scientists for building a collaborative workflow with the aid of LLM agents. Conversely, the expertise can contribute to enhancing LLM agents, as LLMs are essentially trained on text sequences, including dialogues, publications, and computer code. 
As the LLM agents largely remove the barriers of learning and programming,  the existing curriculum of chemical science can include more components of statistical machine learning and data analysis with the assistance of LLM agents.

\section{Case Studies}

\subsection{Physics-Informed Machine Learning for Automated Block Copolymer Phase Identification}

Nature has long mastered the synthesis and use of well-defined macromolecules in biology. While this level of structural specificity remains out of reach with most synthetic polymers, significant progress has been made in preparing precise polymers and developing new strategies to access well-defined materials in high-throughput.\cite{destefano2021biology, barnes2015iterative, jiang2016iterative, al2015synthesis, coin2007solid} When these methods leverage common laboratory equipment that is simple to use and broadly available, it can facilitate widespread use in answering fundamental questions or carefully tailoring structure–property relationships for a specific application.\cite{murphy2024chromatographic} For example, recently Hawker and co-workers have demonstrated the use of automated chromatography to rapidly generate block copolymer libraries.\cite{zhang2020rapid} Block copolymers are an important class of materials that self-assemble into a rich array of nanoscale morphologies.\cite{bates201750th} Key to applications, such as advanced separation membranes, thermoplastic elastomers, photonic crystals, micro-electronics, and drug delivery, is the ability to tune self-assembly through synthetic handles, including block chemistry, block sequence, composition, molecular weight, and dispersity using controlled polymerization techniques.\cite{wang2022ultra, zhang2021biological, maji2022styrenic, moon2020can} This long list of structural variables illustrates the difficulty in navigating and controlling a multidimensional design space. Traditional methods of constructing even an incomplete block copolymer phase diagram involve iterative synthesis followed by multiple purification and isolation steps, which are time-consuming and labor-intensive. The repetitive synthesis of multiple block copolymers is also complicated by slight variations in reaction conditions and/or purification that led to undesired differences among samples and the presence of variable amounts of homopolymer impurities.  

\begin{figure}[t]
    \centering
    \includegraphics[width=\linewidth]{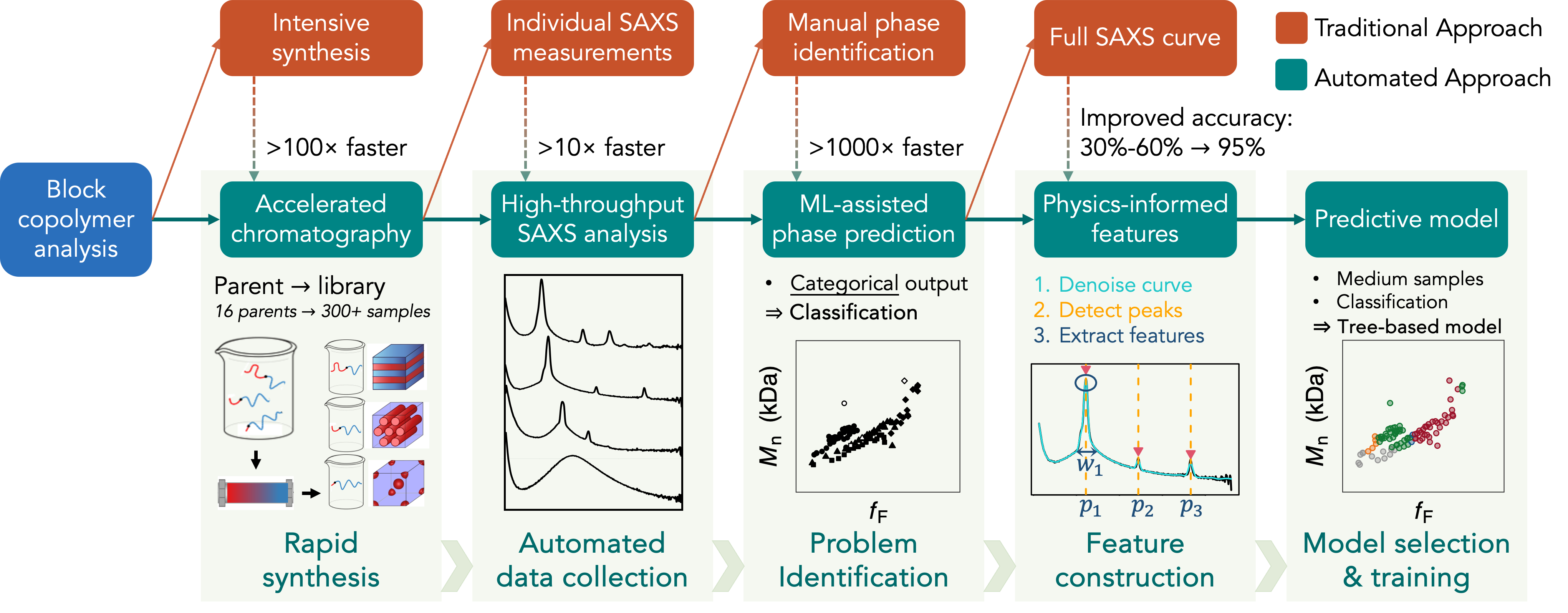}
    \caption{Accelerated workflow for block copolymer phase identification comparing traditional (red) and automated (green) approaches. At each decision point, automated approaches reduce time or improve accuracy compared to conventional methods. 
    }
    \label{fig:workflow_copolymer}
\end{figure}

This process can be substantially accelerated and automated by leveraging the advances of techniques and predictive models shown in Figure \ref{fig:workflow_copolymer}.  For example, a library of 20 well-defined diblock copolymers, spanning a broad range of compositions, was readily prepared in 1 h from a single parent block copolymer and used to prepare an enhanced phase diagram.\cite{zhang2020rapid, murphy2024accelerated, murphy2025high}. 
Because automated chromatography accelerates polymer library construction so significantly, it is essential to pair it with more efficient methods for mapping phase diagrams of diverse block copolymer chemistries. SAXS can determine the polymer phases of these samples, yet it requires an expert to manually identify the phase of the polymer by interpreting SAXS curves, which is time-consuming. This problem was addressed with the development of a physics-informed predictive model to automate polymer phase identification from SAXS\cite{fang2025universal}. Instead of inputting the entire SAXS data into ML models for classifying polymer phases, the authors extend the Kalman filter \cite{kalman1960new} for automated peak detection to extract physics-informed morphological features (PIMF), including the peak locations, width, and sharpness of the peaks. 
These features are used to construct a random forest model,\cite{breiman2001random} 
suitable for classification problems with a small to medium number of training samples. Identifying the phases of hundreds of samples using the random forest model takes less than a second on a desktop computer, and it can be executed without the help of a computational expert.

The  PIMF from SAXS curves  substantially improved the predictive accuracy, achieving around $95\%$ out-of-sample accuracy even for predicting new monomers with different volume fractions not in the database for training  ML models.\cite{fang2025universal} 
The substantial improvement comes from the integration of polymer theory for featurization in machine learning algorithms for determining polymer phases, which dramatically reduces the dimension of the input space in predictions. Furthermore, the maximum prediction probability from a machine learning model, such as a random forest classifier, can be used for quantifying the uncertainty of the prediction. 
The assessed uncertainty enables 
re-inspecting a small subset of the samples with maximum prediction probability lower than a pre-specified threshold, to achieve near $100\%$ accuracy for polymer phase identification. Furthermore, the authors found  3 samples that were mislabeled by the expert but predicted correctly by the ML model.

As polymer phase identification is a new problem for the data scientists, the LLM was used to efficiently acquire domain-specific knowledge about block copolymer behavior and SAXS curves, as illustrated in Figure \ref{fig:LLM_agent_dialogue}(c). This LLM-assisted process accelerates the learning process required in interdisciplinary collaboration. This example  illustrates 
the integration of advanced experimental approaches and data-driven predictive models combined with domain expertise can expedite characterizing structure-property relationships.  

\subsection{ML-Guided Experimental Screening for Discovery of DNA-Stabilized Silver Nanocluster Fluorophores}

DNA-stabilized silver nanoclusters (DNA-Ag\textsubscript{N}) are ultra-small fluorescent nanoparticles with highly tunable properties. First reported in 2004, DNA-Ag\textsubscript{N} contains only 10 to 30 silver atoms stabilized by one to three single-stranded DNA oligomers.\cite{petty2004dna, schultz2012silver,guha2023electron} DNA-Ag\textsubscript{N} are attractive for their sequence-tuned excitation and emission wavelengths that can be tuned from blue to near-infrared (NIR) by the DNA template sequence.\cite{huard2018atomic, swasey2018high} Together with high quantum yields and extinction coefficients, these properties make DNA-Ag\textsubscript{N} promising emitters for biosensing, bioimaging, and nanophotonics.\cite{neaccsu2020unusually, mastracco2023beyond}  For example, emerging NIR-emitting DNA-Ag\textsubscript{N} could enable deep tissue imaging within the NIR tissue transparency window, where biological tissues and fluids are highly transparent to electromagnetic radiation.\cite{hong2017near}

The unique sequence-programmed nature of DNA-Ag\textsubscript{N} presents opportunities to engineer these emitters precisely for specific applications, but DNA-Ag\textsubscript{N} design is highly challenged by the large number of possible templating DNA sequences. Most sequences do not yield useful fluorescent DNA-Ag\textsubscript{N}, and the rules connecting DNA sequence to DNA-Ag\textsubscript{N} properties are complex.\cite{copp2014magic} Moreover, very few X-ray crystal structures of DNA-Ag\textsubscript{N} have been reported, and first-principles computational modeling is currently intractable for DNA-Ag\textsubscript{N} design.\cite{huard2018atomic, sapnik2025structure, cerretani2019crystal,romolini2025shining}

\begin{figure}[t]
    \centering
    \includegraphics[width=\linewidth]{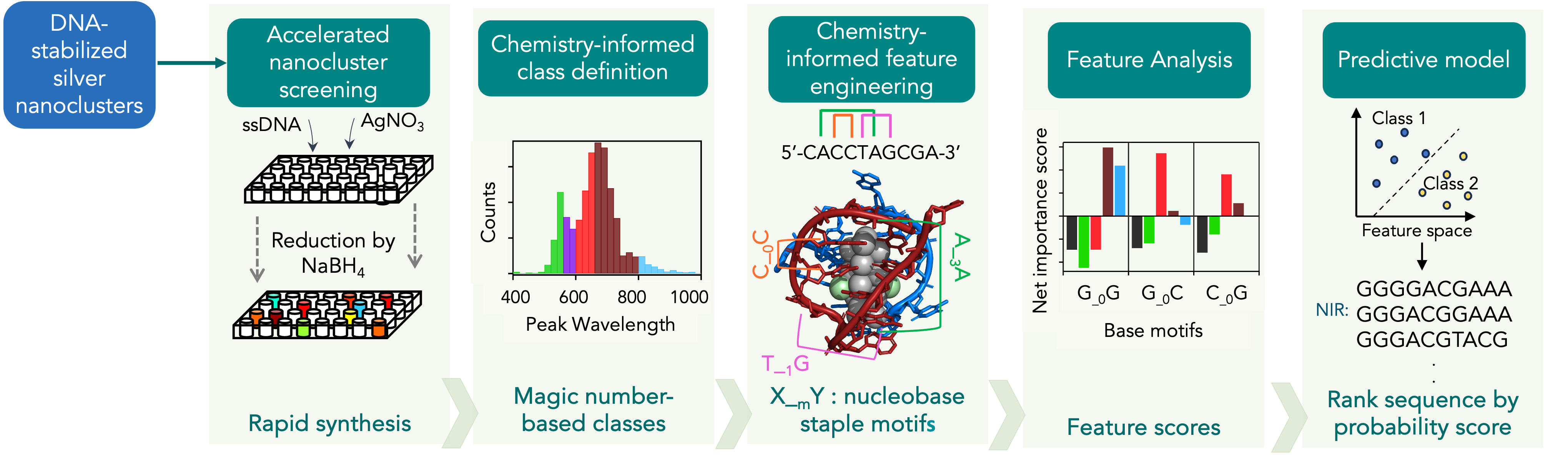}
    \caption{Workflow for ML-enabled DNA-Ag\textsubscript{N} discovery. Experimental DNA-Ag\textsubscript{N} synthesis is performed on 10\textsuperscript{3} DNA oligomers with different sequences, and automated fluorimetry is used to generate training data for ML models. Chemical information guides the choice of the ML problem definition and feature engineering, enabling predictive ML with limited experimental training data and interpretation of sequence-to-property relationships learned by the model.}
    \label{fig:DNA}
\end{figure}

Copp, Bogdanov, and coauthors have developed approaches that combine high-throughput experimental synthesis and characterization with ML models\cite{mastracco2022chemistry, sadeghi2024multi, moomtaheen2022dna,copp2014base,copp2018fluorescence} 
to significantly increase DNA-Ag\textsubscript{N} design efficacy, using the workflow in Figure \ref{fig:DNA}.
First, automated liquid handling is used to synthesize DNA-Ag\textsubscript{N} on 10\textsuperscript{3} different DNA oligomers in well plates, with one oligomer sequence per well. The fluorescence spectrum of each sample is then collected using automated fluorimetry with a well plate reader; universal UV excitation via the nucleobases is employed to excite all DNA-Ag\textsubscript{N} with a single wavelength for rapid fluorimetry. Finally, automated spectral fitting is used to determine the spectral peak parameters for each DNA sequence, thereby generating a large data library that connects DNA sequences to DNA-Ag\textsubscript{N} fluorescence.  

This dataset has been leveraged to train chemistry-informed classification models,  
due to the 
quantized “magic number” properties of nanoclusters, which naturally yield certain DNA-Ag\textsubscript{N} sizes.\cite{copp2014magic} 
Chemically informed featurization has been essential for ML classifiers to learn sequence-to-color relationships, rather than using simple methods such as one-hot encoding.
For example, by featurizing DNA sequence using nucleobase ``staple" motifs inspired by DNA-Ag\textsubscript{N} crystal structure,\cite{ cerretani2019crystal} 
 support vector machines \cite{cortes1995support} were trained to predict the emission color class of a DNA-Ag\textsubscript{N} given input DNA sequence.\cite{mastracco2022chemistry} To ensure robust performance, these models should incorporate regularization techniques and ensemble methods to mitigate overfitting and data imbalance issues commonly encountered in nanocluster datasets. More recently, deep learning models that perform automatic feature extraction and enable continuous property design were introduced and demonstrated for DNA-Ag\textsubscript{N}.\cite{sadeghi2024multi, moomtaheen2022dna}  Beyond prediction, ML models can provide valuable chemical insights into how DNA sequence influences DNA-Ag\textsubscript{N} color through interpretability analysis using feature analysis tools such as BorutaSHAP.\cite{kursa2010feature} 
 
 Experiments have verified the efficacy of ML-guided design approaches for DNA-Ag\textsubscript{N}. One of the most notable findings is the discovery of NIR-emitting DNA-Ag\textsubscript{N}, which are rare in training data libraries, yet can be designed at a 12.3 times enhanced success rate using ML-guided sequence selection.\cite{mastracco2022chemistry} This strategy illustrates the strength of integrating domain knowledge (DNA-Ag\textsubscript{N} chemistry) and ML algorithms to facilitate the systematic discovery of materials and to enhance fundamental chemical understanding in ways that are not achievable using conventional methods.

\subsection{Open-Source Bayesian Optimization Tool for Reaction Development in Small-Molecule Organic Synthesis}

Experimental optimization is ubiquitous in small-molecule organic synthesis. These optimization problems are usually high-dimensional, with reaction spaces defined by both categorical variables (e.g. reagent and solvent identities) and continuous variables (e.g. catalyst loading and temperature). A synthetic chemist selects the initial reaction space to explore based on successful conditions for similar reactions, mechanistic reasoning, and chemical intuition, then iteratively performs rounds of experiments with varied conditions to seek the optimum. The most common conventional strategy for exploration of this space, namely one-variable-at-a-time (OVAT) testing, has proven effective, but is inefficient for exploring a large number of variables and 
overlooks interactions between variables. 

Bayesian optimization (BO)  is well-suited to reaction optimization, as it can suggest multiple experiments by utilizing the quantified uncertainty of a predictive model in a search space defined by both categorical and continuous parameters, to ultimately identify the global optimum in a low-data regime.\cite{shields2021bayesian} 
In 2021, the Doyle group developed Experimental Design via Bayesian Optimization (EDBO), an open-source Python package for reaction development.\cite{shields2021bayesian} The algorithm was tuned using real-world experimental data mined from the chemical literature, with the optimizer offering the best performance using a Gaussian process surrogate model\cite{gardner2018gpytorch} and parallel expected improvement\cite{mockus1975bayes} as an acquisition function. The acquisition function suggests batches of experiments that maximize expected utility until the objective is optimized or the reaction space is explored sufficiently that the probability of finding an improved condition is low. This platform can be used in diverse settings for any parameterizable reaction, including everyday bench-scale experimentation and automated systems, making it widely applicable for modern chemical laboratories. 

To benchmark the EDBO algorithm's performance against the choices of human experts, Doyle and coworkers developed a computer game that asked the player to find the highest-yielding conditions for a Pd-catalyzed C–H arylation reaction within a search space of 1,728 possible reaction conditions, defined by three categorical variables (solvent, ligand, and base identity) and two continuous variables (temperature and concentration). To mimic a real laboratory, the resource budget was limited: players chose 5 experiments to run ``per workday" and had 20 ``workdays" to maximize the yield of the reaction. The experimental outcomes supplied to the players were real, with the yield data for every possible reaction being collected beforehand via HTE. 

\begin{figure}[t]
    \centering
\includegraphics[width=1\linewidth]{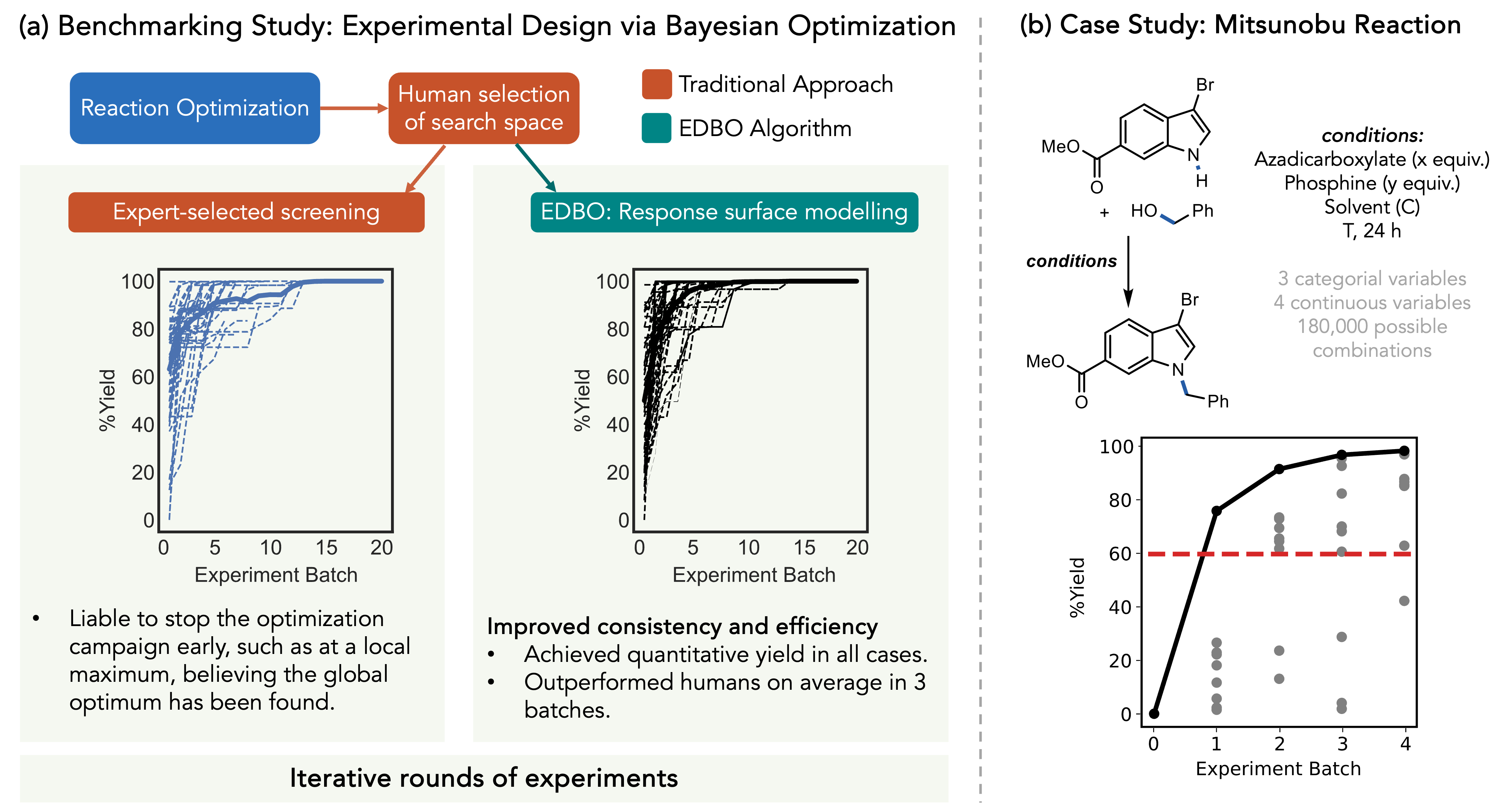} \caption{Experimental Design via Bayesian Optimization. (a) Validation of Bayesian reaction optimization via direct comparison between human performance (left) and machine learning performance (right); optimization curves for individual players or and optimizer runs (dashed) and average (solid) as a function of experiment batch (size: 5). (b) Optimization of a Mitsunobu reaction via EDBO: cumulative best observed yield (black) and individual experiment outcomes (grey) as a function of experiment batch (batch size: 10), yield for standard reaction conditions (red dashed). Adapted with permission from Doyle and coworkers.\cite{shields2021bayesian}}
\label{fig:EDBO}
\end{figure}

For performance comparison, 50 expert chemists were asked to play the benchmarking game and the EDBO algorithm was asked to play it a corresponding 50 times (Figure \ref{fig:EDBO}a). While human experts selected higher-yielding conditions on average for the first round of experiments, the optimizer's average performance surpassed humans' average performance in only three ``workdays" and typically achieved quantitative yield within the first ten. In addition to EDBO's greater efficiency, it displays improved consistency: the optimizer identified the optimal conditions every time it played the game, while many humans participants concluded they had identified the best conditions before achieving quantitative yield and stopped optimization early.

To demonstrate the platform’s ability to optimize real-world reactions used in pharmaceutical development, Doyle and coworkers applied EDBO to a test case of the Mitsunobu reaction.\cite{shields2021bayesian} This reaction was selected because it is used frequently in synthesis, but tends to deliver moderate yields under standard conditions. Methyl 3-bromo-1H-indole-6-carboxylate and benzyl alcohol were chosen as substrates. These substrates afforded a moderate 60\% yield of the desired product under the standard conditions used at Bristol Myers Squibb. Seven total categorical and continuous reaction parameters were selected to define the reaction space: the identity and equivalents of the azadicarboxylate reagent, the identity and equivalents of the phosphine reagent, the identity and concentration of the solvent, and the temperature. Chemical information about the reagents and solvent was encoded in the form of DFT-computed descriptors. With 6 azadicarboxylates, 12 phosphines, 5 equivalencies for each reagent, 5 solvents, 4 concentrations, and 5 temperatures, the full reaction space consists of 180,000 possible combinations. 

With the search space in hand, EDBO was initialized with conditions chosen at random. Ten reactions were run in parallel per experiment batch. The optimizer identified three conditions that delivered the product in nearly quantitative yield (99\%) in only four rounds, totaling 40 experiments (Figure \ref{fig:EDBO}b). EDBO’s ability to deliver a suite of distinct optimized conditions is advantageous, as it enables chemists to choose between several options based on additional factors such as cost and operational convenience.

In 2022, the Doyle group expanded the utility of EDBO with the release of EDBO+.\cite{torres2022multi} The upgraded platform accommodates multi-objective optimization and allows the user to modify the reaction space during the optimization campaign. These improvements adapt the system well to common use-cases in organic synthesis, where multiple objectives (e.g. yield, selectivity, cost) are often in play and condition space is routinely updated as the system is better understood. In addition to its availability as an open-source software package, EDBO+ can be used via a web-based application with a step-by-step graphical user interface designed for users who have little to no coding knowledge, which helps bridge the gap between data scientists and experimental chemists. Furthermore, the integration of EDBO+ as a decision-making tool with other data-driven technologies is already showing promise: the year after its release, EDBO+ proved effective for the optimization of a pyridinium salt synthesis via continuous flow with semi-automated low-resolution data processing,\cite{dunlap2023continuous}, which is gaining popularity for automated reaction development.\cite{sans2015self, cortes2018autonomous}

\section{Summary and Outlook}

Chemical lab research has been transformed by the availability of large volumes of digital data generated by high-throughput experimental facilities that are increasingly automated. These data offer unique opportunities to develop new approaches and algorithms to substantially accelerate the discovery process. 
A key step to advance lab research is to formulate lab tasks as mathematical questions, which is crucial to leveraging progress in machine learning algorithms and AI tools. As many chemical tasks involve identifying unknown relationships, a suitable predictive model can open doors for numerous applications, including accelerating experimental design, processing, and optimization of material properties. To bridge the knowledge gap between distinct areas, LLM agents can help chemical scientists select suitable predictive models, provide standard computer code, and assist computational experts in understanding domain knowledge for developing algorithms to facilitate the discovery process. Furthermore, the answers from LLM agents may inspire new ideas and facilitate the discovery process. Yet LLM agents may generate inaccurate responses and can fabricate or hallucinate information about non-existent theorems or references, which may lead to unsafe experiments, such as providing access to synthesis information that poses security issues. Prompt engineering, including providing contexts and examples, breaking large research questions into smaller pieces, and integrating co-scientists specializing in different domains, can guide LLMs to generate more accurate solutions.\cite{luo2025leveraging} 
Some of these strategies require not only domain knowledge, but also more understanding of data science. Thus, integration of more statistical thinking and machine learning concepts into the pedagogy of chemical science, can assist chemists in better interacting with LLM agents and ensuring the correctness of LLM-derived solutions. 

Overcoming several other common challenges can lead to fruitful outcomes in advancing lab research. First, many experimental characterization tools 
produce data that are, to varying degrees, closed-source, meaning that access to the data is restricted to an ecosystem supported only by the vendor. Recent efforts have been made to facilitate connections between closed-source vendor ecosystems and external software (e.g. LIMS, ELN, or analysis tools) by gaining access to application programming interfaces (APIs) directly from the vendors. For example, a software development kit in a common programming language (Python) was developed and released to consume the API for the Chemspeed instrument, thereby providing greater access to system commands.  
\cite{seifrid2024chemspyd}  Efforts to convert proprietary data into standard formats and share them in an open-source repository can cultivate community efforts. The availability of a standard format of data has driven, for instance, the progress in LLMs and accurate protein structure prediction tools, such as Alphafold.\cite{jumper2021highly}  Furthermore, 
there is a vast need to develop standard software that can be easily plug-in into daily experimental tasks, including automating data processing, making reliable predictions of  chemical relationships,  generating interpretable analysis of experiments, and suggesting solutions for experimental challenges. These tools need to overcome several challenges, including  the limited number of training samples in experiments,  automating model training processes,  enabling uncertainty assessment and assimilation to integrate different types of data. 
 On the other hand, a deeper understanding of the assumptions behind these tools enables chemists to better deploy them in suitable scenarios, identify the reasons when ML tools do not work well, and resolve problems more quickly when interacting with AI agents. 
Together, the joint efforts in experimental and computational fields can substantially accelerate the discovery process 
in chemical science.

\begin{acknowledgement}
This work is supported by 
 BioPACIFIC Materials Innovation Platform of the National Science Foundation
under Award No. DMR-2445868 (NSF BioPACIFIC MIP), and  the  UC Multicampus Research Programs and Initiatives
(MRPI)  under Grant No. M23PL5990.
 We thank Rose Yu from UC San Diego for helpful suggestions.
\end{acknowledgement}



\bibliography{References_2025_edited}

\end{document}